# Food Supply Chain and Business Model Innovation

Saeed Nosratabadi [1], Amirhosein Mosavi [2,3,4*] and Zoltan Lakner [5]

[1] Doctoral School of Management and Business Administration, Szent Istvan University, Godollo 2100, Hungary; saeed.nosratabadi@phd.uni-szie.hu
[2] Faculty of Health, Queensland University of Technology, 130 Victoria Park Road, Queensland 4059, Australia;
[3] Institute of Research and Development, Duy Tan University, Da Nang 550000, Vietnam
[4] Faculty of Humanities and Social Sciences, Oxford Brookes University, Oxford OX3 0BP, UK; a.mosavi@brookes.ac.uk
[5] Department of Food Economics, Faculty of Food Science, Szent Istvan University, Villanyi str. 29-43, 1118 Budapest, Hungary; lakner.zoltan@etk.szie.hu
*Corresponding: a.mosavi@brookes.ac.uk, amirhoseinmosavi@duytan.edu.vn

**Abstract:** This paper investigates the contribution of business model innovations in improvement of food supply chains. Through a systematic literature review, the notable business model innovations in the food industry are identified, surveyed, and evaluated. Findings reveal that the innovations in value proposition, value creation processes, and value delivery processes of business models are the successful strategies proposed in food industry. It is further disclosed that rural female entrepreneurs, social movements, and also urban conditions are the most important driving forces inducing the farmers to reconsider their business models. In addition, the new technologies and environmental factors are the secondary contributors in business model innovation for the food processors. It is concluded that digitalization has disruptively changed the food distributors models. E-commerce models and internet of things are reported as the essential factors imposing the retailers to innovate their business models. Furthermore, the consumption demand and the product quality are two main factors affecting the business models of all the firms operating in the food supply chain regardless of their positions in the chain. The findings of the current study provide an insight into the food industry to design a sustainable business model to bridge the gap between food supply and food demand.

## 1. Introduction

The world population is increasing by 3 billion by 2050 [1] that subsequently lead in an increase in the demand for the food productions. On the other hand, it has been revealed, that that the energy, consumed per person increased from 2250 kcal (9400 KJ) in 1960s to 2880 kcal (12000 KJ) in 2015 [2]. Despite the acceptable performance of global food system in supplying food and decreasing the numbers of undernourished people, one in eight people were suffering from severe undernourishment in 2014 [3] and 815 million people in 2018 [4].

In addition to the demand side, the research shows that the food supply is facing serious problems due to climate changes. Drought, rising temperatures, changes in precipitation regimes, increase of $CO_2$ levels are named the most critical issues influence the yields of agricultural productions, and these issues are expected to exacerbate in the next 50 years [5]. Such changes subsequently result in socio-economic factors such as the increase of the prices [6]. Hence, to meet this steadily increasing food demand, the current food supply chain system and activities should be reconsidered.

The food supply chain (FSC) consists of a chain of activities elaborating how a product is produced and delivered to the final consumers. At each stage of the chain, value or values are added to the product by each player of the FSC (i.e., farmers, processors, distributors, and retailers). Therefore, along with the supply chain, there is a chain so-called value chain explaining the value/values are added to a product in each step. In other words, numerous actors perform in each stage of the FSC to produce the final product from raw material and deliver it to the final consumers. Each of the entities have their objectives which may be contrasted with the other actors' as the activities of each entity influence the performance of the whole supply chain [7]. The concept of business model provides the ability to design and analyze the value a business is offering and delivering to its customers [8]. The business model explains the position of a business in the value chain [9]. All the FSC actors have their own business models and they try to do their best to design elegantly and accurately their business models to increase competitiveness. Moreover, social [10], economic, and environmental factors [11] affect the design of business models of businesses in the food supply chain. Therefore, survival in the FSC is hard to manage [12] and it depends on the uniqueness of the business model.

Hence, analyzing the business model of all FSC players can provide the answers for many questions related to the food supply. Besides, any action to increase the food supply for meeting the future demand for food can be related to the business model of the FSC players. Thus, the main objective of the current study is to provide an insight illustrating how business models and innovations in business models contribute to businesses in different parts of the FSC to bridge the gap between food supply and food demand for future generations. To do so, a systematic literature review conducted to find current solutions that are considered to optimize the production and deliver healthy foods to the consumers. Following, the methodology applied in this study is elaborated in detail, and after that, the findings are provided. It is worth mentioning that to provide a better understanding of the concepts have used in this study, such as the FSC and business model strategies, are defined and explained in advance. Ultimately, the discussion, contributions, and the possible implementation of the findings is provided.

## 2. Food Supply Chain

The food supply chain (FSC) comprises several stages in which food travels from the farmers to the final consumers [1]. In other words, a network of different actors in each stage of the FSC produces and delivers a final product to meet final customers' needs. Much research is conducted to investigate and analyze the FSC, while the general consensus is that the main FSC actors are farmers, processors, distributors, retailer, and consumers (e.g. [7] [85]). In a such FSC, the farmers harvest the initial production, processors produce the final products and packages them, distributors supply the final products to the retailers and finally, the retailers are the ultimate places that consumers purchase the products [1]. To analyze the FSC in the current study, the proposed model of Vorst [86] is admitted. According to Vorst [86], the FSC consists of farmers, food processors, distributors, retailers, and consumers handling.

According to the methodology section, 72 documents constitute the database of the current study. According to table 1, the research objective of 12 out of 72 documents focus on farmers, 21 of them have done a research on the food processors, nine documents concentrates on food distributors, 18 documents analyze the retailers in the food industry, four documents concentrate on the consumption and customer handling activities, and eight documents have targeted the entire FSC. The advantage of classifying the documents based on their focus on the supply chain is that it facilitates further analyses on the business models have been studied in each stage of the FSC.

**Table 1.** Categorizing the reviewed articles based on their focus on the FSC.

| Explanation | Farmer | Processor | Distributor | Retailer | Consumer | The Whole Value chain |
|---|---|---|---|---|---|---|

| Sources | Blasi, et. al [70]; Pölling, et. al. [74]; Krivak, et al. [28]; Hooks, et al. [29]; Morris, et. a. [54]; Panța [21]; Pölling, et. al [58]; Robinson, et. al [60]; Siame [64]; Tushar, et al. [66]; van Eijck, et al. [83]; Varela-Candamio [84]. | Kähkönen [25]; Lange and Meyer [13]; Zucchella and Previtali [14]; Liberti, et al. [15]; Bhaskaran, and Jenkins [30]; Bogers and Jensen [76]; De Bernardi, and Tirabeni [35]; Di Matteo and Cavuta [78]; Giacosa [40]; Harringon and Herzog [42]; Hemphill [43]; Hutchinson, et al. [44]; Jolink, and Niesten [45]; Jia, et al. [80], Markowska, et al. [50]; Mars [51]; Morris, et al. [82]; Sardana [62]; Svensson and Wagner [65]; Vojtovic, et al. [68]. | Samuel, et al. [26]; Shih, and Wang [72]; Wubben, et al. [73]; Berti, et al. [16]; Bruzzone, et al. [32]; Gitler [41]; Hong, et al. [79]; Kim, et al. [19]; Martikainen, et al. [52]. | Di Gregorio [24]; Chang, Wei, and Shih [33]; Dawson [34]; Fiore, et al. [36]; Franceschelli and Santoro [37]; Franceschelli, et al. [38]; Huang, et al. [18]; Karpyn, et al. [46]; Kaur, and Kaur [47]; Lin, L., et al. [48]; Lu, et al. [20]; Massa and Testa [53]; Morris, et al. [55]; Pereira, et al. [57]; Ribeiro, et al. [59]; Russell and Heidkamp [61]; Sebastiani, et al. [63]; Cheah [77]. | Martinovski [71]; Ukolov, et al. [27]; Balcarová, et al. [75]; Franchetti [39]. | Adekunle, et al. [31]; Barth, et al. [17]; Long, et al. [49]; Minarelli, et al. [81]; Pahk and Baek [56]; Vivek [22]; Ulvenblad, et al. [23]; Zondag, et al. [69]. |
|---|---|---|---|---|---|---|
| Numbers | 12 | 21 | 9 | 18 | 4 | 8 |

## 3. Business Model Innovation and Business Model Strategies

The concept of business model provides the opportunity for the entrepreneurs and organizational decision-makers to analyze the logic of their businesses [87]. Indeed, the business model simply explains what values a business creates, to whom, and how it can make money through the value creation and value delivering processes [87]. Many frameworks and models are offered in the literature to analyze a business model, but all the models strive to explain four main aspects of a business: 1) value proposition, which refers to the products and services the business is providing, 2) value delivering, which implies the mechanisms the business is connected with its final customers to deliver the products and services to them, 3) value creation, points out the main activities which are necessary to create and deliver the values to the customers, and 4) value capturing, which indicates the ways a business makes money through the value creation and delivering processes [88].

According to Gambardella and McGahan [89], BMI is the adoption of novel approaches to commercialize underlying assets. In other words, when a BMI happens that value proposition and the business logic are changed. Amit and Zott [90] believe that BMI occurs in three ways: 1) doing the current business and bonding the current activities in new ways, 2) innovation in the ways the current activities execute, and 3) formulating new activities. Many driving forces are mentioned in the literature that induces the businesses to innovate their business model. New inventions, human capital, and new technologies are spelled as the most frequently reasons imposing the businesses to reconsider their business models [91]. BMI is just not a passive response to the environmental changes, but also it has been considered as a strategy to take advantages of the changes and create competitive advantages for the business [92].

Therefore, the firms in the FSC encounter with five strategies to innovate their business model: 1) innovating the value proposition, 2) reconsidering the value delivering mechanisms, 3) innovating the value creation processes, 4) providing new value capturing models, and 5) proposing a quite new business model.

## 4. Methods

A systematic literature review on base of Prizma methods conducted to meet the main objective of the current study. Figure 1 depicts the procedure and the steps taken to generate a database of all the relevant published documents that have focused on business models in FSC. The first step was to identify the maximum number of articles are published in the common field of the business models and the food supply chain. To do so, firstly, the databases of Thomson Reuters Web-of-Science (WoS), Elsevier Scopus, Science Direct, Emerald Insight, J-store, and Sage Publications, which are the most preferred databases for the research related to the economic and the business management disciplines, selected for the further consideration. Then, the search query of business model/business models/business model innovation and food (i.e., TITLE-ABS-KEY ("business model*") AND TITLE-ABS-KEY (food*) and TITLE-ABS-KEY ("business model*") AND TITLE-ABS-KEY (food*)) for the title, abstract, keywords, or source title applied in the mentioned databases to identify the published documents in the common area of these two fields. The terms business model and business models searched separately as some of the databases displayed different results and differentiate these two terms. On the other hand, the word 'food' considered to find the related articles, as most of the articles have not utilized the term of food supply chain directly while they have done a research in one of the FSC stages. The combination of search queries maximized the number of published documents in the common field of the business models and the FSC.

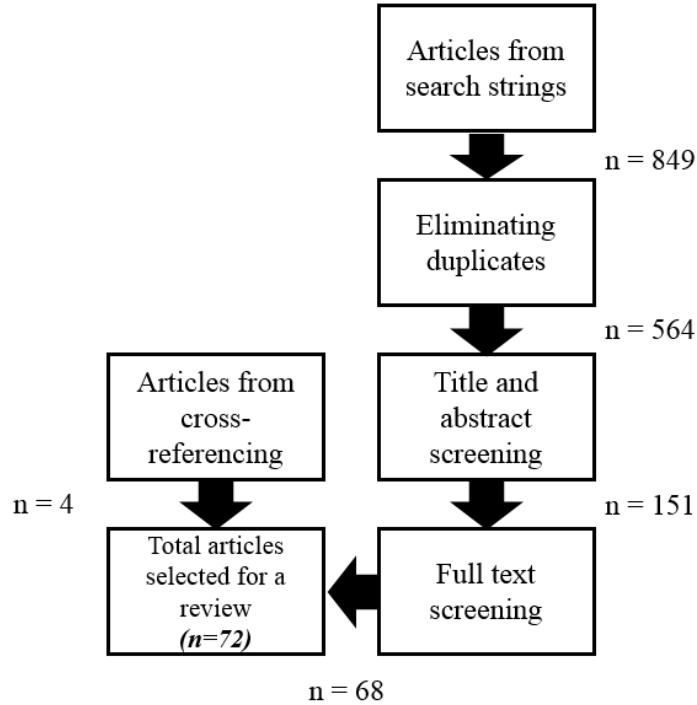

**Figure 1.** Summary diagram of the systematic selection process.

In addition, the search was limited to peer-reviewed journals, conferences, and books/book chapters written in English and published in the period of 1999 to November 2019. As it is presented in figure 1, 849 documents identified as the result of the initial search. Elsevier Scopus and Thomson Reuters Web of Science respectively with 516 and 204 documents (out of 849 found documents) had the most published documents, and the J-Store (with seven documents out of 849 documents) and the Sage publications (with four documents out of 849 documents) had the least share of documents among the other databases considered in this study (see figure 2).

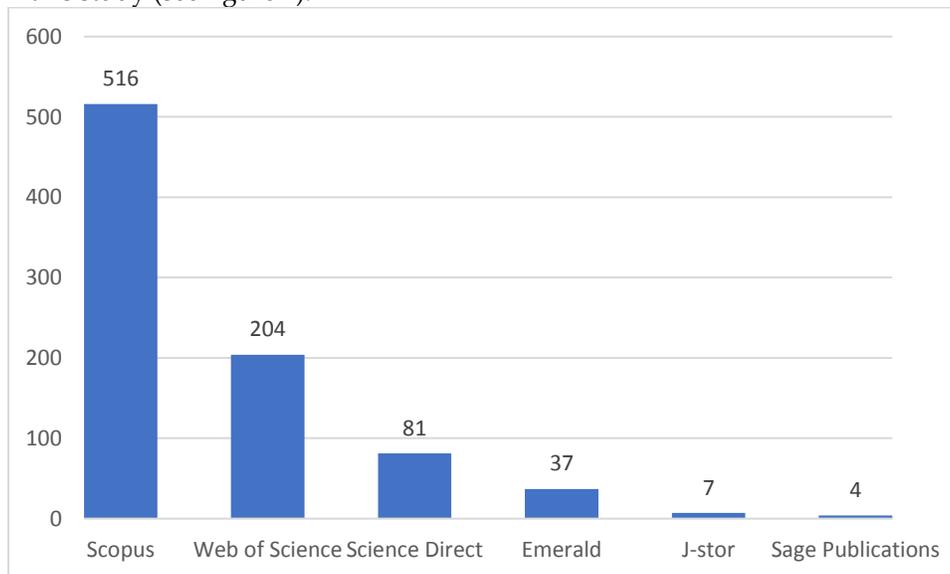

**Figure 2.** Results of initial search string, last update: November 2019.

Articles from the primary search string are categorized by country of origin in which the studies were conducted. It is disclosed that China, as a developing country, has the higest share of the publications on business model innovation (BMI) and FSC. India is also another developing country that has shares of publications in the common area of BMI and FSC. The rest of studies took place in developed countries such as the US, Italy, the UK, France, Netherlands, Australia, Germany, and Spain (see figure 3).

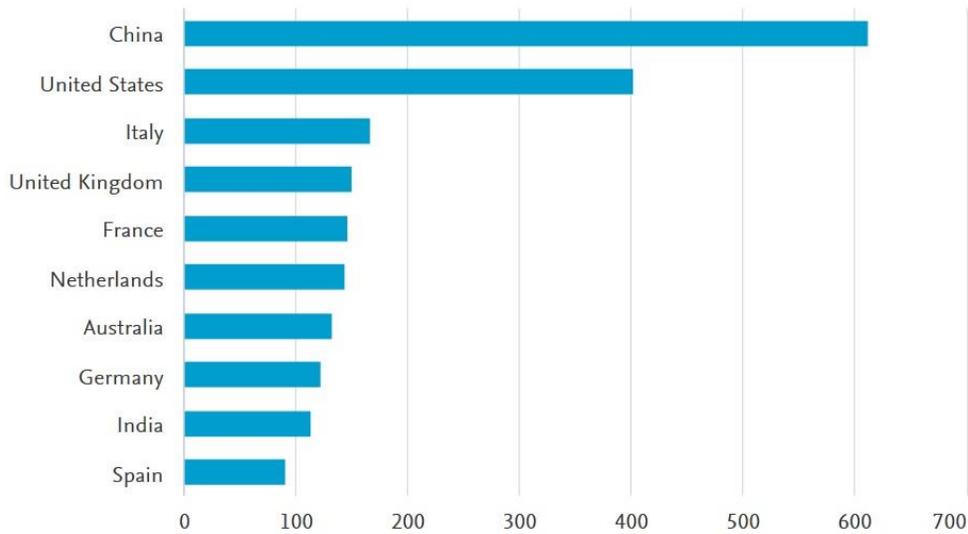

**Figure 3.** Distribution of publications on business models and food supply chain in different countries

According to Figure 1, the second step in selection of final documents was eliminating the duplicated documents which appeared in more than one database during the first step. This step has been carried over independently by a two-member panel of authrors. In case of debate a consensus discussion has been initiatied, involving a third member of authrors' collective. After this step, 564 individual documents had been identified. In the next step, the title and the abstracts of these 564 articles monitored precisely so as to find the relevant documents studied the common area of the BMs and the FSC. The output of this step resulted in 151 documents ready for the next step. To ensure finding the relevant articles, the full text of 151 documents were studied in detail. As a result, 68 documents were selected for the final analysis as they had all the criteria to meet the objective of this study, since all these 68 articles targeted the common research area of the BMs and the FSC. In addition, four more documents found very suitable for further analysis based on cross-reference checking. Therefore, 72 documents considered for the final analysis to investigate how the BMI provides solutions to improve the FSC performance. Figure 4 illustrates the trends of the past two decades of publications in the common area of business models and foods.

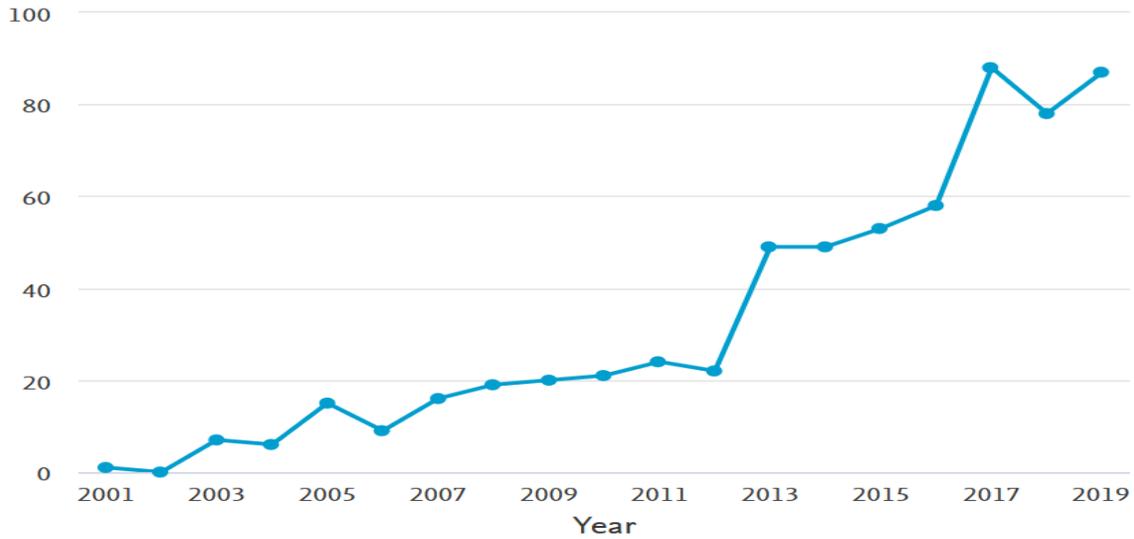

**Figure 4.** Trends of publications in the common area of the business models and the foods from 1999 to 2019.

A close look at the generated database of this study exposed that 53 out of 72 documents were original research articles, 9 of them were conference papers, 5 of them were review articles, 4 of them were book chapters and one of the documents was a commentary, published in a peer-review journal (see figure 5).

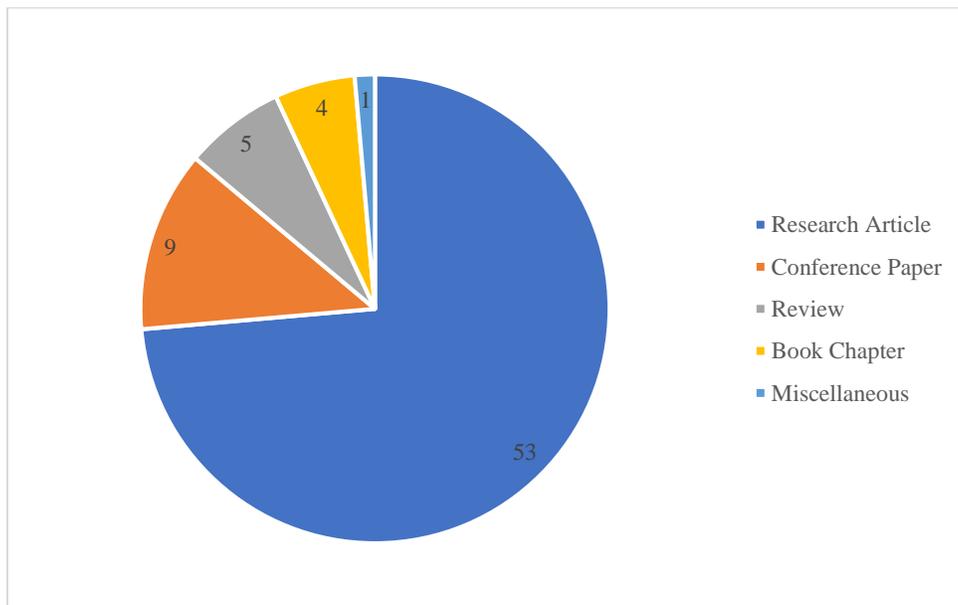

**Figure 5.** Types of documents considered in the current study.

In table 2, the journals having the most share of published documents rowed respectively. British Food Journal with the six documents and Journal of Cleaner Production with four documents are journals published the most articles. According to table 2, British Food Journal, Journal of Cleaner Production, Sustainability, Business Strategy and the Environment, and Journal of Agriculture Food Systems and Community Development published 23% of the documents (17 out of 72 documents).

**Table 2.** Journals with the largest number of documents.

| Journal name | Number of documents |
|---|---|
| British Food Journal | 6 |
| Journal of Cleaner Production | 4 |
| Sustainability | 3 |
| Business Strategy and the Environment | 2 |
| Journal of Agriculture Food Systems and Community Development | 2 |
| Total | 17 |

The vast majority of the documents (46 out of 72) utilized qualitative empirical research to meet their objective. Quantitative empirical research and conceptual papers with respectively 15 and 11 documents were the other research types utilized among the selected documents in this study (see table 3).

**Table 3.** Documents based on research type.

| Explanation | Conceptual | Qualitative Empirical | Quantitative Empirical |
|---|---|---|---|
| Sources | Lange and Meyer [13]; Zucchella and Previtali [14]; Liberti, et al. [15]; Berti, Mulligan, and Yap [16]; Barth, Ulvenblad, and Ulvenblad [17]; Huang, Lee, and Lee [18]; Kim, Lee, and Yang [19]; Lu, et al. [20]; Panța [21]; Soundarrajan and Vivek [22]; Ulvenblad, et al. [23]. | Di Gregorio [24]; Kähkönen [25]; Samuel, Shah, and Sahay [26]; Ukolov, et al. [27]; Krivak, et al. [28]; Hooks, et al. [29]; Bhaskaran, and Jenkins [30]; Adekunle, et al. [31]; Bruzzone, et al. [32]; Chang, Wei, and Shih [33]; Dawson [34]; De Bernardi, and Tirabeni [35]; Fiore, Conte, and Conto [36]; Franceschelli and Santoro [37]; Franceschelli, Santoro, and Candelo [38]; Franchetti [39]; Giacosa, Ferraris, and Monge [40]; Gitler [41]; Harringon and Herzog [42]; Hemphill [43]; Hutchinson, Singh, and Walker [44]; Jolink, and Niesten [45]; Karpyn, and Burton-Laurison [46]; Kaur, and Kaur [47]; Lin, L., et al. [48]; Long, Looijen, and Blok [49]; Markowska, Saemundsson, and Wiklund [50]; Mars [51]; Martikainen, Niemi, and Pekkanen [52]; Massa and Testa [53]; Morris, Jorgenson, and Snellings [54]; Ogawara, Chen, and Zhang [55]; Pahk and Baek [56]; Pereira, et al. [57]; Pölling, et al. [58]; Ribeiro, et al. [59]; | Blasi, Ruini, And Monotti [70]; Martinovski [71]; Shih, and Wang [72]; Wubben, Fondse, and Pascucci [73]; Pölling, Sroka, and Mergenthaler [74]; Balcarová, et al. [75]; Bogers and Jensen [76]; Cheah, Ho, and Li [77]; Di Matteo and Cavuta [78]; Hong, et al. [79]; Jia, et al. [80]; Minarelli, Raggi, and Viaggi [81]; Morris, Shirokova, and Shatalov [82]; van Eijck, et al. [83]; Varela-Candamio, Calvo, and Novo-Corti [84]. |

| | Robinson, Cloutier, and Eakin [60]; Russell and Heidkamp [61]; Sardana [62]; Sebastiani, Montagnini, and Dalli [63]; Siame [64]; Svensson and Wagner [65]; Tushar, et al. [66]; Ulvenblad, Ulvenblad, and Tell [67]; Vojtovic, Navickas, and Gruzauskas, [68]; Zondag, Mueller, and Ferrin [69]. | |
|---|---|---|
| Number | 11 | 46 | 15 |

The data collection method was the other characteristic checked among the documents and it turned out that most articles collected their data from multiple sources. However, 47% of articles used case studies to collect their data. Literature synthesis, questionnaire administration, secondary data, and interview are respectively the other sources of data collection among these 72 articles (see figure 6). It is worth noting that multiple sources and tools, such as participant observation, focus groups and document analysis, interview, and survey, were used to conduct the case study to collect data.

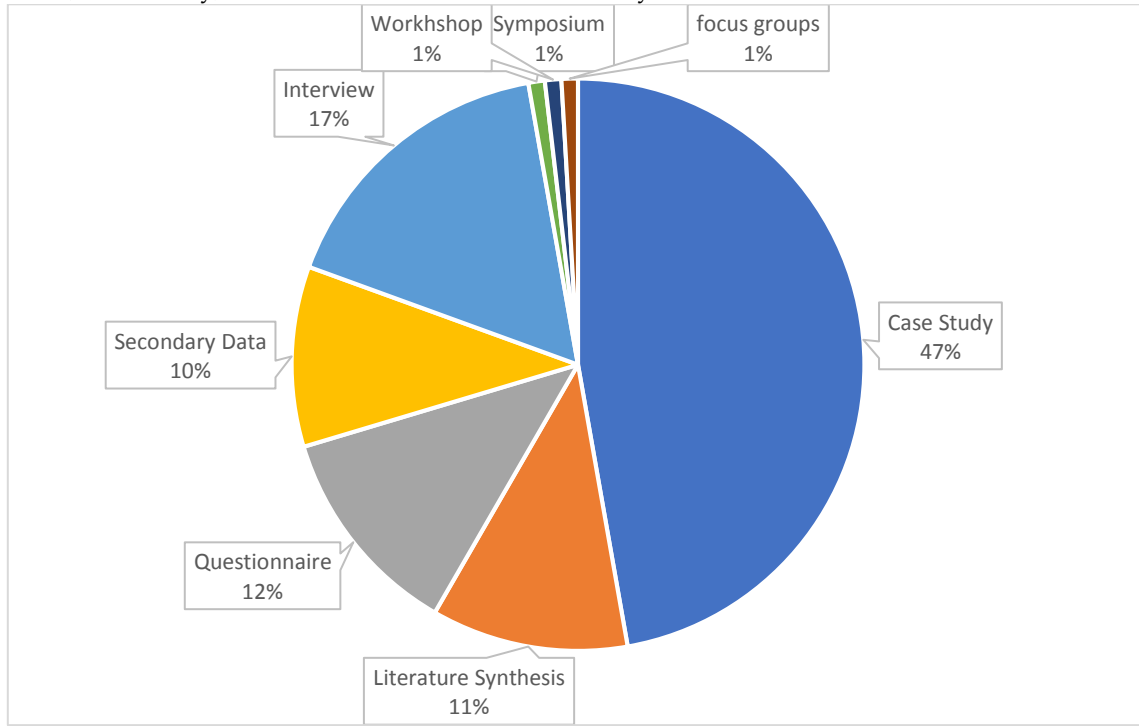

**Figure 6.** Data Collection Method of the documents considered in the current study.

In order to provide a better understanding of the analyses presented in the current study, the concepts of food supply chain, business models innovation, and business model strategies are shortly discussed, and then the finding provided.

## 5. Findings

The finding concluded from reviewing the database of the current study reveals that solutions for improving the business model will vary based on the position of a business in the FSC. To analyze the solutions, the documents categorized based on the part of the FSC they have targeted. In addition, it clarified what business model strategy each paper used to innovate the business model.

*5.1. Business Model Innovation in the Food Supply Chain*

25 out of 74 reviewed documents in the current study have provided solutions to BMI of the firms and entities of the FSC. Table 4 classifies these 25 documents based on their focus on the FSC and the business model strategies. It means that the documents are firstly classified according to which part of the supply chain is focused on the purpose of the article. On the other hand, the position of each document in each row of table 4 reflects the business model strategy that the document applied to BMI. Each of these articles is described below in detail on the basis of their position in the supply chain.

Table 4. Business model innovation and supply chain and Business model strategies.

| Explanation | Farmers | Processors | Distributors | Retailers | Consumers | The entire supply chain | Numbers |
|---|---|---|---|---|---|---|---|
| Value Proposition | | Kähkönen [2] | | Di Gregorio [3] | Martinovski [4] | | 3 |
| Value Creation | Pölling, Sroka, and Mergenthaler [5] | | | Huang, Lee, and Lee [6] | | | 2 |
| Value Delivering | | | Shih and Wang [7], Kim, Lee, and Yang [8] | Kaur and Kaur [9], Pereira et al. [10] | | | 4 |
| Value Capturing | | | | | | | 0 |
| Business Model | Pölling et al. [11], Varela-Candamio, Calvo, Novo-Corti [12] | Liberti et al. [13], Vojtovic, Navickas, and Gruzauskas [14], Giacosa, Ferraris, and Monge [15], Jolink and Niesten [16], | Berti, Mulligan and Yap [24], Martikainen, Niemi, and Pekkanen [17] | Cheah, Ho, and Li [18], Franceschelli, Santoro, and Candelo [19], Ribeiro, Sobral, Peças, Henriques [20], Lu et al. [21] | | Adekunle et al. [22], Barth, Ulvenblad and Ulvenblad [23], Pakh and Baek [24], Ulvenblad et al. [25] | 16 |
| Numbers | 3 | 5 | 4 | 8 | 1 | 4 | 25 |

5.1.1. Farmers

According to table 4, three of the documents proposed solutions to BMI for the farmers in the FSC in which Pölling, Sroka, and Mergenthaler [74] consider innovation in the value proposition as a solution to BMI for the farmers. While, Varela-Candamio, Calvo, Novo-Corti [84], and Pölling et al. [58] provide a new business model for the farmers in the FSC. The following is a summary of these studies.

*Value Creation*

Pölling et al. [74] elaborate the importance of city-adjustment in success of urban farming. They sort a set of strategies such as high-value production, direct marketing, and tourism services and also, they introduce business models such as 'low-cost specialization', 'differentiation', and 'diversification' for adjusting the farms in the urban areas. Research findings by Pölling et al. [74] resulted from the investigation of 180 urban farms in Ruhr Metropolis, Germany discloses that the city-adjusted farms reported a better economic performance and anticipated a more positive prospect compared to the non-city-adjusted farms which did not use the mentioned strategies and business models.

Business Models

Varela-Candamio et al. [84] propose a conceptual framework to design green business models in which rural women play multiple critical roles in generation-production-consumption of functional foods as producer, educator/advisor, and buyer of such products. Where rural women are considered the main educators in the families to boost social-environmental awareness. In the production stage, rural women have a more proactive role in producing functional foods as farmers who are tied with academic institutions so as for transferring the knowledge and producing the functional foods. While, in the consumption stage, the role of rural women constitutes demanding the functional foods. Moreover, Varela-Candamio et al. [84] debate that functional foods comprise elements added naturally or processed, either to increase human health and well-being or mitigate the risk of diseases.

Utilizing case studies, Pölling et al. [58] develop new solutions to design urban farming business models. According to Pölling et al. [58], farming and agriculture in the urban areas requires unique business models which are distinctive from rural areas. They identify three business models for the urban farming called differentiation, diversification, and low-cost specialization. Where the business model of 'differentiation' is associated with niche production and differentiation. Differentiation business model recommends the urban farmers to analyze the whole value chain and utilize vicinity to the final consumers and, by a vertical integration, capture more values. Pölling et al. [58] argue that to perform successfully a differentiation business model, not only the integration is important, but also the product should possess specific features such as exotic species or traditional breeds. Pölling et al. [58] believe that 'diversification' in urban farming business model consists the variety in the value proposition the farmer offers to the customers. They also articulate that agro-tourism, social events (i.e., education, therapy, health), horse services, and care farming are a sort of services that urban farms frequently offer to their clients as well. To justify economically urban farming, since the farmland in and around urban areas are smaller than rural areas, higher added values crop production is necessity. Therefore, Pölling et al. [58] explain that 'low-cost specialization' business model is an urban farming model in which only the products with high added values, high transportation costs, freshness, and high perishability are produced because the vicinity to the final consumers is a competitive advantage and makes the model feasible.

5.1.2. Processors

5 out of 25 articles in table 4 provide solutions for designing and performing the business models for the processors in the FSC where Kähkönen [25] proposes strategies to innovate the value proposition in order to innovate a business model in the food industry and Liberti et al. [15], Giacosa et al. [40], Vojtovic et al. [68], and Jolink and Niesten [45] provide new business models for the processors in the FSC.

Value Proposition

Kähkönen [25] studies the concept of value net in the context of the food industry. Kähkönen [25] defines the value net as "a dynamic, flexible network comprising the relationships between its actors who create value through collaboration by combining their unique and value-adding resources, competences and capabilities". The finding of her study reveals that the value net business model significantly affects the performance of food companies. The finding of the study also illustrates that the actors in the value net in the food industry look for competitive advantages through networking and joint projects. On the other hand, Kähkönen [25] claims that since the main aim of value net is to provide value/values to customers through a collaborative process in which shared knowledge leads to innovative and powerful value proposition for food processors.

Business Model

Liberti et al. [15] work on an EU funded project called i-REXFO. The main objective of i-REXFO is to design a business model which is able to diminish landfilled food wastes through actions reducing food wastes and producing energy from the inevitable wastes. The i-REXFO model includes four phases. The first phase consists providing a database to design a tool to analyze the feasibility of the i-REXFO approach in the desired area. The second phase focuses on strategies to minimizing the expired food in the retailers. Liberti et al. [15] sort a set of strategies to reduce the expired food such as setting strategic prices and communication policies for pre-expiration food, increasing consumer awareness about food expiration label, collecting and distributing unsold pre-expired foods to charities, providing doggy bags among HORECAs. The third phase of i-REXFO business model for avoiding landfilling is to generate energy from expired food in which the food wastes are collected from the retailers and HORECAs and processed to produce biomass biogas plant for electricity production. According to Liberti et al. [15] to test replicability and transferability of the i-REXFO model, this model will be performed in Spain and Hungary.

Giacosa et al. [40] conduct a case study to investigate the approaches to strengthening the business models of family food businesses. They realize that tradition, the family's values and experiences in the food sector, and innovation, the creation of new values and opportunities by injecting new ideas, are two main pillars strengthen the business model of a family food business. Giacosa et al. [40] explain that utilizing the customers' feedback results in the product innovation, and it also presents the opportunity to increase the quality and product ranges and subsequently, it will lead to the customer satisfaction. This is because they can offer a wider range of traditional products in old and new flavors utilizing both traditional approaches and the modern technologies. Giacosa et al. [40] provide evidences revealing that considering the tradition and innovation in the family food businesses improve and affect not only the processes of value proposition and value creation but also the models of value delivering and capturing. Vojtovic et al. [68] provide an innovative framework to design a sustainable business model for the processors in the food and beverage industry. Inspired by the business model canvas, Vojtovic et al. [68] propose a ten-pillar business model to develop a sustainable business model for the processors in the food and beverage industry.

They suggest that in the first step, the business concept should be explained where key principles and values that the business offers to the costumers, sustainable benefits to the society and the environment, and the company vision and long-term goals should be clearly identified. After explanation of the business concept, the second step is to identify the customers. Vojtovic et al. [68] divide the customers to three categories of early adopters, niche market and mass segment. The third pillar of this model is building relationships, including branding, habit-forming, and legislation issues. Designing a distribution channel is the next pillar of the proposed sustainable business model of Vojtovic et al. [68]. Vojtovic et al. [68] articulate that planning for resources, designing the key activities (i.e., operating, support, and development) to run the business model, and developing a sophisticated support system are respectively fifth, sixth, and seventh pillar of this model. In this model, developing the partners network with suppliers, manufacturers, service providers is the next action. Estimating the cost structure and selecting the income model are the last two pillars of this model. Jolink and Niesten [45], utilizing the concept of Ecopreneurship, try to develop sustainable business models for the organic food industry. According to Jolink and Niesten [45], ecopreneurs are subcategory of sustainable entrepreneurs where the business operates in the mass markets and tries to meet the sustainability goals (i.e., economics', environment's, and society's benefits) at the same time. The result of their study exposes four ecopreneur business models among the organic food companies. The income business model, the subsistence business model, the growth business model, the speculative model.

Jolink and Niesten [45] argue that the income business model is adopted by small companies whose axial objective is to generate income through creating the opportunity to consumers to eat healthy foods. Providing the proper information to the consumers about eco-products plays a critical role in this model. In accordance with the findings of Jolink and Niesten [45], the objective of the companies applying subsistence business model "…is to survive and meet basic financial obligations". Although they try to make the world better, being ecologically sustainable is not in their priority, since they need to reach the mass markets, lack of sufficient organic raw materials restricts them to present eco-product to their customers. Therefore, they need to make a compromise between being economically and environmentally sustainable. The third ecopreneur business model discovered among the organic food companies is the growth model, where the focal point is to invest and reinvest on the financial aspects and the relationship with the customers in order to be profitable in the long term. According to Jolink and Niesten [45], those companies implement such a business model have a relatively large impact on the market. These companies have turned being sustainable to a competitive advantage and have become profitable in this way. Jolink and Niesten [45] express that the speculative model is the fourth ecopreneur business model they identified among the organic food companies. According to Jolink and Niesten [45], the speculative model focuses on making money by selling eco-products where the economic profits set in the priority. Indeed, in this model, sustainability turned into a tool for profitability. These ecopreneurs concentrate on short-term goals

5.1.3. Distributors

Among the research documents reviewed in the current study, four of them study business models of food distributors in which Shih and Wang [72] and Kim et al. [19] investigate new solutions for delivering the food productions to the customers and Berti et al. [16], and Martikainen et al. [52] introduce new business models for the distributors in the FSC.

Value Delivering

One of the most important issues in the FSC is food distribution, where cold chain management plays a vital role. Having a frozen storage with the risk of high-energy consumption and cool storage with the threat of bacterial decay is a dilemma the distributors in the food industry deal with. Hence, Shih and Wang [72] by means of an Internet-of-Things (IoT) architecture and ISO 22,000, an international food standards, propose four solutions to overcome the aforementioned problems in the food distribution: cold chain home delivery service, convenience store (CVS) indirect delivery, CVS direct delivery, and flight kitchen service. According to their results, applying the above-mentioned business models could result in a 1.36 million increase in annual sales of braised pork rice, generating extra revenue of US$6.35 million by creating new distribution channels, and also reducing 10% energy consumption.

Shih and Wang [72] elaborate that cold chain home-delivery service refers to free home delivery of the foods in 1–2 working days for orders exceeding a minimum purchase requirement at the off-peak hours (14:00–17:00). This approach not only provide the opportunity to use less the cold storage but also expands brand recognition and facilitates market penetration. On the other hand, Shih and Wang [72] express that

CVS indirect delivery refers to delivering fresh foods products that are processed in OEM facilities by CVS companies. Convenience store companies prefer cool storage products than the products needed to be thawed where reheating them does not takes more than 30–40 s in a microwave. In addition, Shih and Wang [72] argue that CVS direct delivery refers to the food products are processed, packed, and delivered by CVS. This approach is selected in the case the food quality and food safety are very important. According to Shih and Wang [72], the flight kitchen business model is quite similar to CVS indirect delivery business model where the only difference is the lower supply volume and fewer supply spots. In accordance with the flight kitchen business model, semi-processed food products are delivered to international catering companies via cool storage. Then they process and deliver it to the airplane flight kitchen, where they just need to re-heating it. In this approach, daily delivery is very important to maintain the food safety and quality.

To solve the urban agriculture's problems, Kim et al. [19] propose the Eco-M business model where organic fresh foods produced by suburban agriculture delivered daily to the local markets. Kim et al. [19] claim that although this model has performed successfully, it cannot benefit from competitive price since the risk of wasted food is high as the products are fresh foods and their expiration date is too close to the production date, and they should be consumed in 10 days after production.

Business Model

Berti et al. [16] propose a disruptive business model producing new values and new markets by redefining the food supply chain. They introduce a digital food hub, an online marketplace, facilitating efficient connections among local food producers and consumers. Berti et al. [16] argue that it is a sustainable business model as it increases the demand for the local food, and it also promotes healthy and sustainable food for the local communities. Berti et al. [16] believe that this digital food hub, indeed, provides a strategic network across the food supply chain to co-produce socio-environmental shared values.

Martikainen et al. [52] try to design business models for third-party logistics service provider (LSP) for local food supply chains. Utilizing business model canvas, they propose two new business models named business model for the focused service offering and business model for the full-service offering. The main difference between the focused and full service offerings is the market that they have targeted. In the focused business model, the concentration is on upstream producers and processors, while the full offering covers downstream operators' needs. Therefore, this difference in the target market reflects on a different value proposition and, subsequently, a different business model. Table 5 provides the opportunity to compare two business models by elaborating the models in detail.

**Table 5.** The difference between focused and full-service offering business models.

| Business model components | Focused service offering | Full-service offering |
|---|---|---|
| Value Proposition | We improve our customers' ability to fulfill the service needs of their customers and cut their costs. | We bring together the consumption and the production of local food in a cost-effective and business opportunity providing manner. |
| Customer Segments | Food producers<br>Food processors | Food producers<br>Food processors<br>The customer relationships extend to downstream partner of the supply chain to achieve value-adding service.<br>Retail stores<br>Institutional kitchens<br>Wholesalers<br>Farm product shops<br>Food clubs |
| Customer Relationships | Automated daily routines<br>Personal service (phone) for exceptions and changes | Automated daily routines<br>Personal service (phone) for exceptions and changes.<br>Networking the supply chain stakeholders, feedback channel to supply-side |
| Channels | Personal contacts, solution-seeking approach<br>Long term contracts, service level agreements | Personal contacts, solution, and synergy seeking approach<br>Long term contracts, service level agreements |
| Key Activities | Arrangement of pickups, collecting, and deliveries with supplier and delivery sides<br>Order mediation<br>Invoicing and payment mediation<br>Exceptions and changes management | Arrangement of pickups, storage, collecting and deliveries<br>Inventory and terminal management<br>Invoicing and payment mediation<br>Sales and marketing the customers' products<br>Product quality management<br>Network management |
| Key Resources | Food handling & transportation knowledge<br>Transportation planning knowledge | Food handling & transportation knowledge<br>Transportation planning knowledge<br>Terminal facilities |
| Key Partners | Logistic operators (transportation companies)<br>Accounting and invoicing partner<br>ICT provider | Logistic operators (transportation companies)<br>Accounting and invoicing partner<br>ICT provider<br>Professional reseller and/or marketer of local food |
| Cost Structure | On cost-driven side – minimizing supply chain costs<br>Aiming at economies of scale together with logistics operators´ own volumes (cost-cutting)<br>Avoidance of fixed costs, transaction-based agreements with the partners | On value-driven side – maximizing the value for the supply chain customers<br>Aim to economies of scale together with logistics operators´ volumes.<br>Avoidance of fixed costs |
| Revenue Streams | Contract-based fee<br>Transaction-based pricing preferred<br>Revenues depend on volumes<br>Compensation on flexibility and volume increase | Contract-based fee<br>Priced individually<br>Revenues can be tied to the revenues increase of the customers |

Source: Own construction based on Martikainen et al. [52].

5.1.4. Retailers

The Retailers play a very important role in the FSC as these are the places that the food products are delivered to the final customers. This part of the FSC attracted more researchers as 8 out 25 documents have focused on different aspects of the business model of the retailers. Di Gregorio [24] propose strategies for the value proposition the retailers are delivering, Huang et al. [18] focus on how retailers can create values, Kaur and Kaur [47] and Pereira et al. [57] study innovative solutions for retailers to innovate their business model by redesigning value delivering models. Finally, Cheah, Ho, and Li [77], Franceschelli et al. [38], Ribeiro et al. [59], and Lu et al. [20] identify new business models for the retailers in the FSC. The Following is a summary of all these studies.

Value Proposition

Di Gregorio [24] proposes a creative business model for retailers, where the products are delivered to the customers, in the food industry. He applies the concept of placed-based business model to introduce a model in which location-specific resources are used to create and capture value. Di Gregorio [24] conduct case studies within slow food in Italy (Coop Italia and Eataly). According to his results, the placed-based business model in the slow food industry in Italy will subsequently lead in resilience, sustainability, and prosperity of the social context by reviving passion for traditional food cultures and increasing supply and demand for local food products.

Value Creation

Huang et al. [18] develop a e-business model for food souvenir industry in Taiwan, inspired by e-commerce business model. The focus of their innovation is value creation. According to their model, the final customers are able to order online, and local providers are responsible for the supply and delivery of orders.

Value Delivering

Utilizing a sensor-based measurement containers (SBMCs), an Android application, and cloud IoT-enabled grocery management system (CE-GMS), Kaur and Kaur [47] provide a creative solution to business model innovation for retailers in the FSC. By designing an innovative approach to get the order and deliver it to the customers, they have created a new business model. According to proposed model of Kaur and Kaur [47], when the retailers get the order from a customer, he subsequently get an alarm related to the quantity of the product in the store and the warehouses at the same time. This contribution helps them to manage the quantity of the products and make them able to maximize their potential to cover the customers' needs.

Pereira et al. [57] run a case study to investigate a sustainable business model for delivering fresh milk. Pereira et al. [57] compare traditional channels and vending machines for supplying the fresh milk. Their finding discloses that utilizing vending machines shorten supply chain; therefore, it has a lower impact on the environment due to the elimination of mediators and transportations activities. Pereira et al. [57] also realized that the success of vending machines remarkably depends on consumer behavior. As their finding exposes when the consumption of environment-friendly products is very important to the consumers, the vending machines were more profitable.

Business Model

Cheah et al. [77] provide empirical evidences that business model innovation provides competitive advantages to the retailers in the food industries. Their finding illustrates that the retailers acting in a high turbulence environment have a higher chance to get sustainable competitive advantages by re-innovating of their business model. Franceschelli et al. [38] propose a framework to design a sustainable business model for a food startup in which, in addition to the economic profit, the social and environmental benefits are considered. Their study focused on a pizzeria startup in Italy.

Ribeiro et al. [59] strive to test the sustainability of a retailing strategy so-called ugly business model in which the waste from fresh fruits and vegetables that are not sold through the conventional distribution channels due to the appearance of these products, is minimized. According to their results, this project, in addition to the economic benefits, has had social benefits (i.e., "*increasing waste awareness and healthy food consumption and community engagement in reduction of the waste*", etc.) and environmental benefits (i.e., "*prevent food wastes and climate change mitigation benefits*").

Lu et al. [20] provide solutions for designing a business model for sustainable agricultural products utilizing internet of things (IoT). Aided by IoT, the new business model provides products through networks and e-commerce via electronic data interchange and e-mail online sales contract along with the traditional marketing channels. The convenience of online shopping and instant messaging interoperability are mentioned of the new value propositions that IoT can offer for the sustainable agriculture. Lu et al. [20] also claim that IoT designs a sophisticated information system in the organizations which arms the businesses to design a customer-centric structure collecting the data and customers' feedbacks and also provides the adequate information to the customers.

5.1.5. Consumers

Consumers are the most important part of the FSC, as, without the consumers, all the supply chain will be meaningless. Handling customers' issues and studying their behavior is of the utmost importance during managing the FSC. Martinovski [71] has an innovative approach to design value propositions based on the customers' behavior.

Value Proposition

Martinovski [71] has a different perspective to design a business model for an entity performing in the FSC. Martinovski [71] believes that consumer behavior is the key determinant in designing a business model. Therefore, he proposes the concept of modeling a business model according to consumer behaviors while purchasing food products. His finding reveals that this approach is a tool for the decision-makers to design a sustainable business model in which on the one hand, the businesses are able to utilize this customers centric approach to get the customers' feedbacks so as for developing corresponding value propositions for their target market and on the other hand, the society's and customers' benefits are considered and healthy safe food productions are delivered to them based on their feedback.

5.1.6. The Entire Supply Chain

In addition to the studies that targeted a specific stage of the FSC, many studies have focused on the whole supply chain and have provided solutions to create values for the whole FSC. Designing solutions for value creation and value delivering for the entire supply chain, indeed, implies that the researchers have tried to provide innovative solutions that affects the whole supply chain from the farmers to the retailers and customers. For instance, Adekunle et al. [31] and Barth et al. [17], Ulvenblad et al. [23] and Pakh and Baek [56] recommend frameworks to innovate the business models in the FSC.

Business Model

Adekunle et al. [31] design a business model for small millets value chain in India. According to their result, a mixed CI–PL business model is appropriate for small millets value chain in India where CI-business model refers to customer intimacy business models in which the customer is placed in the center of the business model and PL business model points out to the product leadership business models in which the quality of the product is of the utmost importance. According to the proposed business model of Adekunle et al. [31], there should be an interactive collaboration among farmers, technologists, processors, and researchers to produce and deliver high-quality small millets through innovation, creating and sharing knowledge. This collaboration will lead in increase yield, improve marketing, and reduction of drudgery. Barth et al. [17] develop an approach to design an innovative sustainable business model for the businesses performing in the agri-food sector statements. Based on a deep literature review, the design questions for development of each pillar of the business model. According to Barth et al. [17] the business model constitutes four main pillars of 1) value proposition, 2) value creation and delivery, 3) value capture, and 4) value intention. In table 6 their proposed business model is presented.

**Table 6.** Proposed business model of Barth et al. [23] for the agri-food sector.

| Pillars | Degree of Innovation | Sustainability |
|---|---|---|
| Value proposition | Offers 'more of the same' or something new to the firm/world? Existing markets or new markets? | Do the product/service, customer segments, and relationships enhance sustainability? For example, traceability for products and Standards for safety and quality? |
| Value creation and delivery | Improvements of existing channels or new relationships? Familiar (fixed) networks or new (dynamic) networks (e.g., alliances, joint ventures)? Improvements of existing technologies or new, emerging technologies? | Do key activities, resources, channels, partners, and technologies focus on sustainability aspects? Awareness of food-related ethics? Ethical consumption? For example, ecological sustainability, social justice, and animal welfare. |
| Value capture | Incremental cost-cutting in existing processes or new processes that generate revenues? | Do cost structures and revenue streams include sustainability considerations? For example, sustainable food systems based on environmental, social, and economic aspects. |
| Value intention | Attitudes to change and innovation | Is sustainability a means, a goal, or something else? Is sustainability enhancing or limiting the BM? |

Source: Barth et al. [17].

Ulvenblad et al. [23] study the barriers to business model innovation in the agri-food industry. To do so, they run a systematic literature review where they reviewed 570 research articles published between 1990 to 2014. They, ultimately, categorize the barriers to BMI in the agri-food industry to two classes of internal barriers and external barriers. Where the internal barriers to BMI include: 1) individual barriers (e.g., perceptions, values, behavior), 2) organizational barriers (e.g., lack of competencies, insufficient resources, and unsupportive organizational structure). On the other hand, Ulvenblad et al. [23] articulate that the external barriers to BMI comprise 1) resistance and lack of support from specific actors and 2) restrictive macro-environment. It is worth mentioning that they provide another layer of analysis for these barriers, and for each of the mentioned barriers, they provide sub-variables.

Pakh and Baek [56] develop the concept of 'considerate design approach' to design a sustainable business model in which value propositions have considered to meet the benefits of all the stakeholders. Pakh and Baek [56] develop four business models includes: 1) neighboring producer community, a collaboration platform between the local farmers/producers and the customers for direct sale, 2) local food café, a mediator between local farmers/producers and the customers where the local foods are served, 3) farm mentoring institute, a mentoring platform transferring the farmers' knowledge to the others and students, and finally 4) food community, including cuisine researchers and educators training the locals to utilize local ingredients to cook professionally in order for either their own consumption or selling their foods.

## 6. Discussion

To increase food supply, many solutions are provided in the literature in which 72 published documents have studied the business models of the businesses in the FSC. A deep analysis of these documents illustrates that 25 out of 72 documents present strategies and solutions to innovate the business models in the FSC so that improving the performance of the FSC. Three of these articles provide recommendations to redesign the value propositions in the business models. Where Kähkönen [25] introduces the concept of value net, which suggest collaboration among the different stakeholders to shape the value. Martinovski [71] also has a similar recommendation to design the value proposition. He recommends that to engage the customers in the value shaping processes. While, Di Gregorio [24] has an innovative solution for the value proposition as he considers the location and the place as a source of value.

Whist two of the documents, including Pölling et al. [74] and Huang et al. [18] consider innovation in value creation processes as the strategy to BMI for the food industry. Pölling et al. [74] sort out solutions to adjust the urban farms according to the cities' constraints. On the other hand, Huang et al. [18] provide empirical evidences proving that applying e-commerce models facilitates the value creation processes. Reconsidering the value delivering processes is another strategy are considered by the author to BMI in the FSC. Shih and Wang [72] and Kaur and Kaur [47] recommend applying IoT to optimize the management of delivering the food production. Pereira et al. [57] offer vending machines to delivering the fresh milk product in the urban areas and Kim et al. [19] introduce the concept of Eco-M business model to facilitate delivering fresh foods to the urban areas (see figure 7). These strategies are summarized in figure 6 as the strategies are applied in the FSC to innovate the business model.

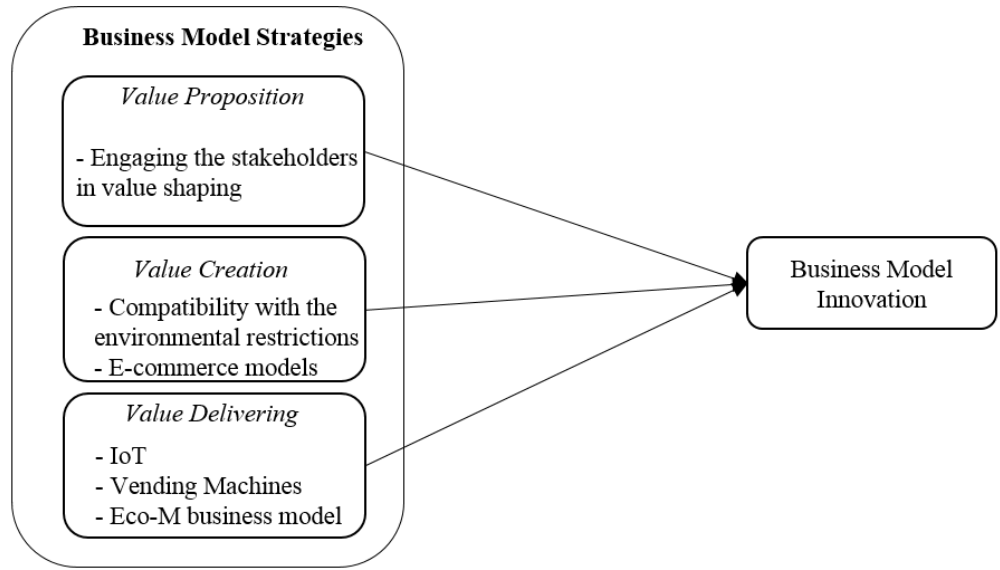

**Figure 7.** Business model strategies to business model innovation in the FSC based on the literature.

Along with the mentioned studies, there are studies present BMI driving forces either for a specific part of the FSC or for the whole of the FSC. For instance, Varela-Candamio et al. [84] claim that women not only play vital roles in designing and implementing a sustainable business models in the farms but also raise the demand for the sustainable products by increasing awareness of local food in communities. Pölling et al. [58] argue that urban farms should adapt their business model according to the conditions of the city. Giacosa et al. [40] see adding innovations and technologies to traditional mechanisms as a source of innovation in business models for the food processors as proposed in [93,94]. Besides, Liberti et al. [15], inspired by the circular economy concept and circular business models, provide recommendations to produce energy from the inevitable wastes. Berti et al. [16] disrupt the current FSC, affected by digitalization. They introduce a digital food hub, which is an online marketplace, to connect the local food producers and consumers. Ultimately, Adekunle et al. [31] consider the customers intimacy and the product quality as driving sources to BMI for the whole FSC (see figure 8).

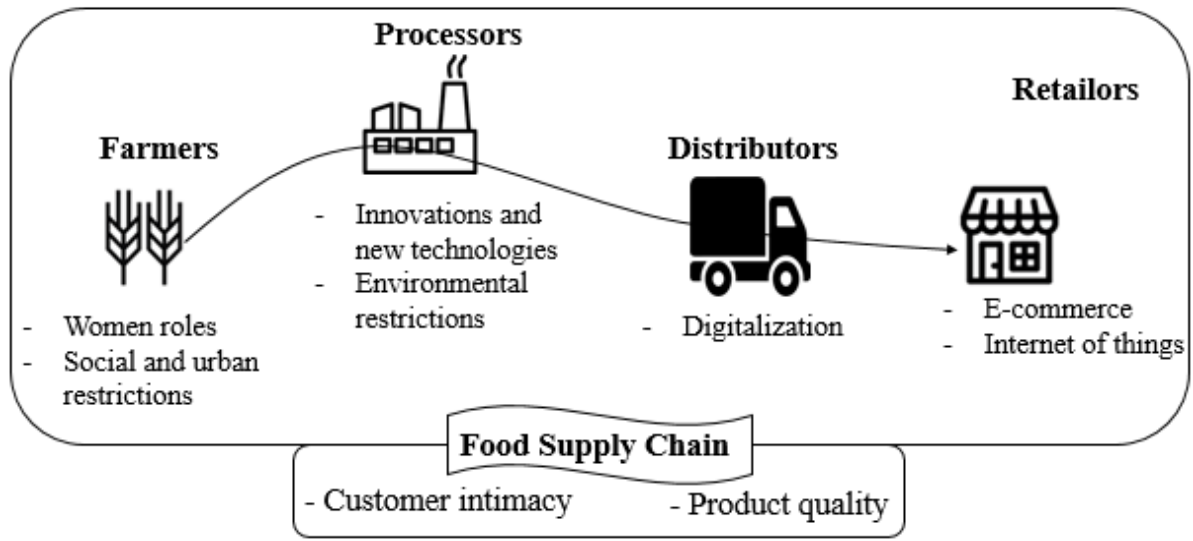

**Figure 8.** The business model innovation driving forces in the food supply chain.

## 7. Conclusion

Food security is a very important issue for both researchers and practitioners struggling to provide solutions for supplying adequate foods to the next generations. Various remedies and recommendations are given in the literature to bridge the gap between food supply and food demand for the next 50 years, including BMI. BMI is a tool allowing the firms in the FSC to optimize values they are creating and delivering to their customers. By means of a systematic literature review, the current study unfolded that the strategies such as engaging the stakeholders in value creation processes, compatibility with the social and environmental constraints, utilizing e-commerce models, and ultimately, applying IoT, vending machines, and Eco-business models for delivering products to customers are considered by the literature to BMI in the FSC. In addition, one of the contributions of the current study was to distinguish the driving forces of BMI based on the position of the firms in the FSC where the literature illustrates that rural women and social and urban conditions are the most important driving forces inducing the farmers to reconsider their business model. Besides, it is disclosed that inventions and new technologies, and environmental issues are the main driving forces to BMI for the food processors. Digitalization has disruptively changed the food distributors models. E-commerce models and Internet of things have been the factors imposing the retailers to innovate their business models. It was also found that customer' needs and product quality are two main factors affecting the business model of all the firms operating in the FSC regardless their position in the chain. At the same time the behaviours of consumers is changing radicall, too. Furher investigations necessitate a complex approach, focusing on consumer-producer interacritons, taking into consideations the changing material and digital environment.

On the one hand, the findings of the current study provide a fundamental insight to the mangers and the entrepreneurs of the food industry to design a business model which has the ability to predict and adapt to environmental changes and also is able to improve the performance of the FSC. On the other hand, these findings can be a basis for the future research. It is recommended to the researchers to design business models for each of the players of the food supply chain based on the findings of this study.

**Author Contributions:** conceptualization, S.N.; investigation, S.N.; writing—original draft preparation, S.N..; methodology, A.M.; writing—review and editing, A.M. and Z.L.; visualization, S.N; validation, ZL.; supervision and controlling the results, Z.L.;

**Acknowledgments:** Financial support of Hungarian Academy of Sciences is acknowledged.

**Conflicts of Interest:** The authors declare no conflict of interest.

**References**

1. Wunderlich, S.M. and N.M. Martinez, Conserving natural resources through food loss reduction: Production and consumption stages of the food supply chain. International Soil and Water Conservation Research, 2018. 6(4): p. 331-339.
2. Pardey, P.G., et al., A bounds analysis of world food futures: Global agriculture through to 2050. Australian Journal of Agricultural and Resource Economics, 2014. 58(4): p. 571-589.
3. Keating, B.A., et al., Food wedges: framing the global food demand and supply challenge towards 2050. Global Food Security, 2014. 3(3-4): p. 125-132.
4. Nations, U. United Nations, Goal 2: end hunger, achieve food security and improved nutrition and promote sustainable agriculture. 2018 [cited 2019 Augus 3rd]; Available from: https://www.un.org/sustainabledevelopment/hunger/.
5. Parry, M.L., et al., Effects of climate change on global food production under SRES emissions and socio-economic scenarios. Global environmental change, 2004. 14(1): p. 53-67.
6. Popp, J., et al., The socio-economic force field of the creation of short food supply chains in Europe. Journal of Food & Nutrition Research, 2019. 58(1).
7. Higgins, A., et al., Challenges of operations research practice in agricultural value chains. Journal of the Operational Research Society, 2010. 61(6): p. 964-973.
8. Nosratabadi, S., et al., Sustainable business models: A review. Sustainability, 2019. 11(6): p. 1663.
9. Mosleh, A. and S. Nosratabadi, Impact of Information Technology on Tehran's Tourism Agencies' Business Model's Components. International Journal of Business and Management, 2015. 10(2): p. 107.
10. Busby, J.S., B.S. Onggo, and Y. Liu, Agent-based computational modelling of social risk responses. European Journal of Operational Research, 2016. 251(3): p. 1029-1042.
11. Borodin, V., et al., Handling uncertainty in agricultural supply chain management: A state of the art. European Journal of Operational Research, 2016. 254(2): p. 348-359.
12. Ahumada, O. and J.R. Villalobos, Application of planning models in the agri-food supply chain: A review. European journal of Operational research, 2009. 196(1): p. 1-20.
13. Lange, L. and A.S. Meyer, Potentials and possible safety issues of using biorefinery products in food value chains. Trends in Food Science & Technology, 2018.
14. Zucchella, A. and P. Previtali, Circular business models for sustainable development: A "waste is food" restorative ecosystem. Business Strategy and the Environment, 2019. 28(2): p. 274-285.
15. Liberti, F., et al., i-REXFO LIFE: an innovative business model to reduce food waste. Energy Procedia, 2018. 148: p. 439-446.
16. Berti, G., C. Mulligan, and H. Yap, diGital food hubs as disruptive business models based on Coopetition and "shared value" for sustainability in the agri-food sector, in Global Opportunities for Entrepreneurial Growth: Coopetition and Knowledge Dynamics within and across Firms. 2017, Emerald Publishing Limited. p. 415-438.
17. Barth, H., P.-O. Ulvenblad, and P. Ulvenblad, Towards a conceptual framework of sustainable business model innovation in the agri-food sector: a systematic literature review. Sustainability, 2017. 9(9): p. 1620.

18. Huang, T.C., T.J. Lee, and K.H. Lee, Innovative e-commerce model for food tourism products. International Journal of Tourism Research, 2009. 11(6): p. 595-600.

19. Kim, J.-B., H.-H. Lee, and H.-C. Yang, Proposal of Eco-M Business Model: Specialty Store of Eco-friendly Agricultural Products Joined with Suburban Agriculture. The Journal of Asian Finance, Economics and Business (JAFEB), 2014. 1(4): p. 15-21.

20. Lu, Y., et al. Research on the innovation of strategic business model in green agricultural products based on Internet of things (IOT). in 2010 2nd International Conference on E-business and Information System Security. 2010. IEEE.

21. Panța, N.D., Arguments in Favor of Moving to a Sustainable Business Model in the Apiary Industry. Studies in Business and Economics, 2017. 12(3): p. 159-170.

22. Soundarrajan, P. and N. Vivek, A study on the agricultural value chain financing in India. Agricultural Economics, 2015. 61(1): p. 31-38.

23. Ulvenblad, P., et al., Barriers to business model innovation in the agri-food industry: A systematic literature review. Outlook on Agriculture, 2018. 47(4): p. 308-314.

24. Di Gregorio, D., Place-based business models for resilient local economies: Cases from Italian slow food, agritourism and the albergo diffuso. Journal of Enterprising Communities: People and Places in the Global Economy, 2017. 11(1): p. 113-128.

25. Kähkönen, A.-K., Value net–a new business model for the food industry? British Food Journal, 2012. 114(5): p. 681-701.

26. Samuel, M.V., M. Shah, and B. Sahay, An insight into agri-food supply chains: a review. International Journal of Value Chain Management, 2012. 6(2): p. 115-143.

27. Ukolov, V., et al., Food-sharing economy pattern comparison in UK and Russian markets. Int. Bus. Manag, 2016. 10: p. 4268-4282.

28. Krivak, G.E., et al., A Mixed-Method Multiple Case Study of Three Business Models for Local Healthy Food Delivery Systems in Underprivileged Urban Areas. 2017.

29. Hooks, T., et al., A Co-operative Business Approach in a Values-Based Supply Chain: A case study of a beef co-operative. Journal of Co-operative Organization and Management, 2017. 5(2): p. 65-72.

30. Bhaskaran, S. and H. Jenkins, Case study of processing firm-distributor firm outsourcing alliance. Journal of Manufacturing Technology Management, 2009. 20(6): p. 834-852.

31. Adekunle, A., et al., Helping agribusinesses—Small millets value chain—To grow in India. Agriculture, 2018. 8(3): p. 44.

32. Bruzzone, A.G., et al., Simulation Based Design of Innovative Quick Response Processes in Cloud Supply Chain Management for "Slow Food" Distribution, in Theory, Methodology, Tools and Applications for Modeling and Simulation of Complex Systems. 2016, Springer. p. 25-34.

33. Chang, L.-H., F.H. Wei, and C.-C. Shih. Sustainable business model for organic agriculture-Lee Zen organic corporation in Taiwan. in III International Symposium on Improving the Performance of Supply Chains in the Transitional Economies 895. 2010.

34. Dawson, J., Changes in food retailing and their implications for new product development, in Consumer-Driven Innovation in Food and Personal Care Products. 2010, Elsevier. p. 25-52.

35. De Bernardi, P. and L. Tirabeni, Alternative food networks: Sustainable business models for anti-consumption food cultures. British Food Journal, 2018. 120(8): p. 1776-1791.

36. Fiore, M., A. Conte, and F. Conto, RETAILERS TOWARDS ZERO-WASTE: A WALKTHROUGH SURVEY IN ITALY. Italian Journal of Food Science, 2015.

37. Franceschelli, M.V. and G. Santoro. SUSTAINABLE BUSINESS MODEL INNOVATION: AN OPPORTUNITY FOR THE FOOD INDUSTRY. in 10th Annual Conference of the EuroMed Academy of Business. 2017.

38. Franceschelli, M.V., G. Santoro, and E. Candelo, Business model innovation for sustainability: a food start-up case study. British Food Journal, 2018. 120(10): p. 2483-2494.

39. Franchetti, M., Development of a novel food waste collection kiosk and waste-to-energy business model. Resources, 2016. 5(3): p. 26.

40. Giacosa, E., A. Ferraris, and F. Monge, How to strengthen the business model of an Italian family food business. British Food Journal, 2017. 119(11): p. 2309-2324.

41. Gitler, J., et al., Social Enterprise Business Sustainability of the Food Banking Model The Case of Leket Israel, Israel's National Food Bank. BUSINESS PEACE AND SUSTAINABLE DEVELOPMENT, 2017(9): p. 76-93.

42. Harringon, R. and C. Herzog, Chef John Folse: a case study of vision, leadership & sustainability. Journal of Hospitality & Tourism Education, 2007. 19(3): p. 5-10.

43. Hemphill, T.A., The global food industry and "creative capitalism": The partners in food solutions sustainable business model. Business and Society Review, 2013. 118(4): p. 489-511.

44. Hutchinson, D., J. Singh, and K. Walker, An assessment of the early stages of a sustainable business model in the Canadian fast food industry. European Business Review, 2012. 24(6): p. 519-531.

45. Jolink, A. and E. Niesten, Sustainable development and business models of entrepreneurs in the organic food industry. Business Strategy and the Environment, 2015. 24(6): p. 386-401.

46. Karpyn, A. and H. Burton-Laurison, Rethinking research: Creating a practice-based agenda for sustainable small-scale healthy food retail. Journal of Agriculture, Food Systems, and Community Development, 2013. 3(4): p. 139–143-139–143.

47. Kaur, J. and P.D. Kaur, CE-GMS: A cloud IoT-enabled grocery management system. Electronic Commerce Research and Applications, 2018. 28: p. 63-72.

48. Lin, L., et al. Market Research and Suggestion of O2O Business Model To "ZHOU HEI YA". 2016. Thirteenth International Conference on Innovation and Management (ICIM2016).

49. Long, T.B., A. Looijen, and V. Blok, Critical success factors for the transition to business models for sustainability in the food and beverage industry in the Netherlands. Journal of cleaner production, 2018. 175: p. 82-95.

50. Markowska, M., R.J. Saemundsson, and J. Wiklund, Contextualizing business model developments in Nordic rural gourmet restaurants, in Handbook of Research on Entrepreneurship in Agriculture and Rural Development, G.A. Alsos, et al., Editors. 2011, Edward Elgar: Cheltenham UK. p. 162-179.

51. Mars, M., From bread we build community: Entrepreneurial leadership and the co-creation of local food businesses and systems. Journal of agriculture, food systems, and community development, 2015. 5(3): p. 63-77.

52. Martikainen, A., P. Niemi, and P. Pekkanen, Developing a service offering for a logistical service provider—Case of local food supply chain. International Journal of Production Economics, 2014. 157: p. 318-326.

53. Massa, S. and S. Testa, Beyond the conventional-specialty dichotomy in food retailing business models: An Italian case study. Journal of Retailing and Consumer Services, 2011. 18(5): p. 476-482.

54. Morris, C., W. Jorgenson, and S. Snellings, Carbon and energy life-cycle assessment for five agricultural anaerobic digesters in Massachusetts on small dairy farms. International Food and Agribusiness Management Review, 2010. 13(1030-2016-82868): p. 121-128.

55. Ogawara, S., J.C. Chen, and Q. Zhang, Internet grocery business in Japan: current business models and future trends. Industrial Management & Data Systems, 2003. 103(9): p. 727-735.

56. Pahk, Y. and J. Baek, Stakeholder centred approach to sustainable design: A case study of co-designing community enterprises for local food production and consumption. 2015.

57. Pereira, Á., et al., Fresh milk supply through vending machines: Consumption patterns and associated environmental impacts. Sustainable Production and Consumption, 2018. 15: p. 119-130.

58. Pölling, B., et al., Business models in urban farming: A comparative analysis of case studies from Spain, Italy and Germany. Moravian Geographical Reports, 2017. 25(3): p. 166-180.

59. Ribeiro, I., et al., A sustainable business model to fight food waste. Journal of cleaner production, 2018. 177: p. 262-275.

60. Robinson, C., S. Cloutier, and H. Eakin, Examining the business case and models for sustainable multifunctional edible landscaping enterprises in the phoenix metro area. Sustainability, 2017. 9(12): p. 2307.

61. Russell, S.E. and C.P. Heidkamp, 'Food desertification': The loss of a major supermarket in New Haven, Connecticut. Applied Geography, 2011. 31(4): p. 1197-1209.

62. Sardana, G., Social business and Grameen Danone foods limited. Society and Business Review, 2013. 8(2): p. 119-133.

63. Sebastiani, R., F. Montagnini, and D. Dalli, Ethical consumption and new business models in the food industry. Evidence from the Eataly case. Journal of business ethics, 2013. 114(3): p. 473-488.

64. Siame, M. Social venturing and co-operative entrepreneurship business model (SVCE-bm) for growing MSMEs in Zambia. in European Conference on Innovation and Entrepreneurship. 2016. Academic Conferences International Limited.

65. Svensson, G. and B. Wagner, A process directed towards sustainable business operations and a model for improving the GWP-footprint (CO2e) on earth. Management of Environmental Quality: An International Journal, 2011. 22(4): p. 451-462.

66. Tushar, H., et al., Employing a Transformative Learning Process for Promoting Sustainable Business Model through Organic Agriculture: A Case Study of the Sampran Riverside. St. Theresa Journal of Humanities and Social Sciences, 2018. 4(2): p. 67-92.

67. Ulvenblad, P.-o., P. Ulvenblad, and J. Tell, An overview of sustainable business models for innovation in Swedish agri-food production. Journal of Integrative Environmental Sciences, 2019. 16(1): p. 1-22.

68. Vojtovic, S., V. Navickas, and V. Gruzauskas, Sustainable business development process: the case of the food and beverage industry. Zeszyty Naukowe Politechniki Poznańskiej. Organizacja i Zarządzanie, 2016.

69. Zondag, M.M., E.F. Mueller, and B.G. Ferrin, The application of value nets in food supply chains: A multiple case study. Scandinavian Journal of Management, 2017. 33(4): p. 199-212.

70. BLASI, E., L. RUINI, and C. MONOTTI, Technologies and new business models to increase sustainability in agro-food value chain–Promote quality and reduce environmental footprint in durum wheat cultivation processes. Agro Food Industry Hi-Tech, 2017. 28(6): p. 52-55.

71. Martinovski, S., NUTRITION BUSINESS MODELS OF CONSUMER BEHAVIOUR WHEN PURCHASING SELF-EXPLANATORY FOOD PRODUCTS. Journal of Hygienic Engineering and Design, 2016.

72. Shih, C.-W. and C.-H. Wang, Integrating wireless sensor networks with statistical quality control to develop a cold chain system in food industries. Computer Standards & Interfaces, 2016. 45: p. 62-78.

73. Wubben, E., M. Fondse, and S. Pascucci, The importance of stakeholder-initiatives for business models in short food supply chains: the case of the Netherlands. Journal on Chain and Network science, 2013. 13(2): p. 139-149.

74. Pölling, B., W. Sroka, and M. Mergenthaler, Success of urban farming's city-adjustments and business models—Findings from a survey among farmers in Ruhr Metropolis, Germany. Land use policy, 2017. 69: p. 372-385.

75. Balcarová, T., et al. Farmers Market: Customer Relationship. in Agrarian Perspectives XXV. Global and European Challenges for Food Production, Agribusiness and the Rural Economy, Proceedings of the 25th International Scientific Conference, 14-16 September 2016, Prague, Czech Republic. 2016. Czech University of Life Sciences Prague, Faculty of Economics and Management.

76. Bogers, M. and J.D. Jensen, Open for business? An integrative framework and empirical assessment for business model innovation in the gastronomic sector. British Food Journal, 2017. 119(11): p. 2325-2339.
77. Cheah, S., Y.-P. Ho, and S. Li, Business Model Innovation for Sustainable Performance in Retail and Hospitality Industries. Sustainability, 2018. 10(11): p. 3952.
78. Di Matteo, D. and G. Cavuta, Enogastronomic Tourism: can it mitigate the Intangibility of the Destination? Streetfood as a new Business Model for the Management of Tourist Regions. Procedia Economics and Finance, 2016. 39: p. 347-356.
79. Hong, I.-H., et al., An RFID application in the food supply chain: A case study of convenience stores in Taiwan. Journal of food engineering, 2011. 106(2): p. 119-126.
80. Jia, F., et al., Investigating the feasibility of supply chain-centric business models in 3D chocolate printing: A simulation study. Technological Forecasting and Social Change, 2016. 102: p. 202-213.
81. Minarelli, F., M. Raggi, and D. Viaggi, Innovation in European food SMEs: determinants and links between types. Bio-based and Applied Economics Journal, 2015. 4(1050-2016-85767): p. 33-53.
82. Morris, M.H., G. Shirokova, and A. Shatalov, The Business Model and Firm Performance: The Case of Russian Food Service Ventures. Journal of Small Business Management, 2013. 51(1): p. 46-65.
83. van Eijck, J., et al., Comparative analysis of key socio-economic and environmental impacts of smallholder and plantation based jatropha biofuel production systems in Tanzania. Biomass and Bioenergy, 2014. 61: p. 25-45.
84. Varela-Candamio, L., N. Calvo, and I. Novo-Corti, The role of public subsidies for efficiency and environmental adaptation of farming: A multi-layered business model based on functional foods and rural women. Journal of cleaner production, 2018. 183: p. 555-565.
85. Plà, L.M., D.L. Sandars, and A.J. Higgins, A perspective on operational research prospects for agriculture. Journal of the Operational Research Society, 2014. 65(7): p. 1078-1089.
86. van der Vorst, J.G., Effective food supply chains: generating, modelling and evaluating supply chain scenarios. 2000.
87. Zott, C., R. Amit, and L. Massa, The business model: recent developments and future research. Journal of management, 2011. 37(4): p. 1019-1042.
88. Chesbrough, H., Business model innovation: opportunities and barriers. Long range planning, 2010. 43(2-3): p. 354-363.
89. Gambardella, A. and A.M. McGahan, Business-model innovation: General purpose technologies and their implications for industry structure. Long range planning, 2010. 43(2-3): p. 262-271.
90. Amit, R. and C. Zott, Creating value through business model innovation. 2012, 2012.
91. Grabowska, M., Innovativeness in business models. Procedia Computer Science, 2015. 65: p. 1023-1030.
92. Mészáros, K. and S. Nosratabadi, Business model innovation for transition to the organic agriculture Acta AVADA, 2018. 5: p. 5-15.
93. Saxena, N.; Sarkar, B.; Singh, S.R. Selection of remanufacturing/production cycles with an alternative market: A perspective on waste management. Journal of Cleaner Production 2020, 245, 118935.
94. Ullah, M.; Sarkar, B. Recovery-channel selection in a hybrid manufacturing-remanufacturing production model with RFID and product quality. International Journal of Production Economics 2020, 219, 360-374.